\documentclass[aps,pra,floats,epsfig,amsmath,amsthm,showpacs]{revtex4}

\usepackage{graphicx}

\begin{document}
\title{Hydrogen atom in phase space. The Kirkwood-Rihaczek representation}
\author{L. Praxmeyer,$^1$ and K. W\'odkiewicz$^{1,2}$}
\affiliation{$^1$Institute of Theoretical Physics, Warsaw
University,
Ho\.za 69, 00--681 Warsaw, Poland\\
$^2$Department of Physics and Astronomy,
    University of New Mexico,
    800 Yale Blvd. NE,
    Albuquerque, NM 87131   USA  }

\date{16 December 2002 }

\begin{abstract}
 We present a phase-space representation of the hydrogen atom
using the Kirkwood-Rikaczek distribution function. This
distribution allows us to obtain analytical results, which is
quite unique because an exact analytical form of the Wigner
functions corresponding to the atom states is not known. We show
how the Kirkwood-Rihaczek distribution reflects properties of the
hydrogen atom wave functions in position and momentum
representations.
\end{abstract}


\maketitle

\section{Introduction}

The Wigner function \cite{Wigner}, which was introduced in 1932,
is up to now the most commonly used phase space quasi-distribution
 representation of physical systems in    momentum and position
 representations.
 This function fully characterizes the quantum state and gives basic
physical intuition about the investigated state. Moreover, there
are some simple physical properties that make this function unique
\cite{O'Connell}. Nevertheless, over the years many others
quasi-distributions have been introduced and studied. Especially
well-known are functions from  the Glauber and Cahill
s-parameterized class of distributions \cite{GlaCah} containing
the Wigner function, the Glauber-Sudarshan $P$-representation
\cite{Gla,Sud}, and the Husimi or the $Q$-representation
\cite{Husimi}.

For many systems the Wigner function  is   well known and
 has been carefully studied, but there are some elementary
quantum mechanical states, for which this phase space
representation is not known. An example of such system is the
hydrogen atom. No exact analytical formula of the Wigner function
is known even for the   1s state. In the literature one can
find a limited number of papers devoted to this subject
\cite{Dahl1, Dahl2, Dahl3, Nouri}, and all the published results have been
achieved   using methods of   approximation.

It is the purpose of this short communication to present an
analytical  phase space representation of the hydrogen atom using
a lesser-known Kirkwood--Rihaczek (K-R) distribution function
\cite{Kirkwood}. Its form allows  us to obtain analytical
results, and as we have shown in \cite{K-R} this distribution,
like  the Wigner function, fully characterizes a quantum
state.

\section{Different quasi-distribution and operator ordering}

In general, the problem of defining  a joint probability
function in phase space lies in the  operator ordering. Because
different orderings of position $\hat{q}$ and momentum $\hat{p}$
operators are not equivalent,   the association of the
quantum operator with classical functions  has no unique solution.
The association of classical phase space functions ${\cal A}(q,p)$
with quantum operators  is shown by:
\begin{eqnarray}
 \{ \hat{\cal A}(\hat{q},\hat{p})\}_{\rm ordering}
  =\iint dq  dp\,
{\cal A}(q,p)\,\{ \delta(q-\hat{q})\delta(p-\hat{p}) \}_{\rm
ordering}. \nonumber
\end{eqnarray}
The Fourier decomposition of the Dirac delta functions   makes
it possible to rewrite the above equation  in  the form
\begin{eqnarray}
 \{\delta(q-\hat{q})\delta(p-\hat{p}) \}_{\rm ordering}
=\frac{1}{(2\pi\hbar)^2}
        \iint dq' dp'
        e^{\frac{i(p'q-q'p)}{\hbar}}
        \left\{ e^{\frac{i\hat{p}q'}{\hbar} }
        e^{-\frac{ip'\hat{q}}{\hbar} } \right\}_{\rm ordering}.
\label{dirac}
\end{eqnarray}
For every operator ordering there exists a corresponding
probability quasi-distribution in phase space.

The Wigner distribution function is obtained by taking a quantum
average of the above formula with the respect to the Wigner-Weyl
ordering:
\begin{eqnarray}
W(q,p)=
 \langle \Psi\mid\{\delta(q-\hat{q})\delta(p-\hat{p})
 \}_{WW}\mid\Psi\rangle
=\frac{1}{(2\pi\hbar)^2}
        \iint dq' dp'
        e^{\frac{i(p'q-q'p)}{\hbar}}
       \langle \Psi\mid
       e^{\frac{i(  \hat{p}q'-p'\hat{q} )}{\hbar}}
        \mid\Psi\rangle .
 \label{WW}
\end{eqnarray}
in which $\hat{q} $ and $ \hat{p}$ operators are put in the same
exponent.

The K--R distribution function corresponds to an ordering called
the anti-standard ordering, that is obtained by putting all
$\hat{p} $ operators on the left of all $ \hat{q}$ operators:
\begin{eqnarray}
K(q,p)=
 \langle \Psi\mid \delta(p-\hat{p}) \delta(q-\hat{q})
 \mid\Psi\rangle=
 \frac{1}{(2\pi\hbar)^2}
        \iint dq' dp'
        e^{\frac{i(p'q-q'p)}{\hbar}}
       \langle \Psi\mid
       e^{\frac{i\hat{p}q'}{\hbar} }
        e^{-\frac{ip'\hat{q}}{\hbar} }
         \mid\Psi\rangle
       .
\end{eqnarray}
The complex conjugation of
the K--R function
corresponds to the standard ordering (all operators $ \hat{q}$ are on the left
followed by $ \hat{p}$ operators), and the real part of the K--R distribution
function is obtained by a symmetric superposition of the anti-standard and
the standard ordering:
$\frac{1}{2}(\hat{q}^{n}\hat{p}^{m}+\hat{p}^{m}\hat{q}^{n})$.

Obviously, the Wigner-Weyl ordering and the anti-standard ordering
are not equivalent, so they lead to different quasi-distributions.
But both of these orderings are intuitive and simple ones, and
appropriate phase space distribution functions have  the same
marginal properties.

\section{The Wigner and the
Kirkwood-Rihaczek distribution functions}

The definition of the famous Wigner  quasi-distribution
function was firstly presented as:
\begin{equation}
W_{\Psi}(\mbox{q},p)= \frac{1}{2\pi\hbar}\;\int
\Psi^{\star}(q+\xi/2)\; e^{\frac{ip\xi}{\hbar}}\;\Psi(q-\xi/2)
d\xi\;. \label{zwig}
\end{equation}
Although for most one dimensional systems formula (\ref{zwig})
allows to evaluate the Wigner function relatively  easily,
in three dimensions this problem is more complicated. Integrals
become quite cumbersome and sometimes impossible to handle
analytically. As we have mentioned before,    an analytical
formula   for the Wigner function is not known even for
1s state of hydrogen atom.

That is why, in order to  investigate the hydrogen atom in phase
space we shall use the K--R distribution. This
function was introduced by Kirkwood \cite{Kirkwood} just one  year
after Wigner introduced his function. Then, in 1968, the same
function was rediscovered by Rihaczek \cite{Rihaczek} in the
context of signal time-frequency distributions and is known by his
name in signal transmitting theory.  The definition of the K--R
function in terms of anti-standard ordering has a simple form:
\begin{eqnarray}
K(q,p)&=& \frac{1}{2\pi\hbar} \int d\xi\, \Psi^{\star}(\xi)\,
 e^{\frac{i(\xi-q)p}{\hbar}}\,\Psi(q)=
\nonumber\\
 &=&\frac{1}{2\pi\hbar}\, \Psi(q)\, e^{- \frac{ipq}{\hbar}}\,
\widetilde{\Psi}^{\star}(p). \label{zrih}
\end{eqnarray}
Elsewhere \cite{K-R} we have presented  an extensive analysis of
the K--R distribution  and its comparison with the Wigner
distribution function. Here we will only mention the main
properties of  the K--R function.   Similar to the Wigner
function,  the K-R distribution has the correct marginal
properties:
\begin{eqnarray}
 \int  K(q,p)\,dp\, =\, |\Psi(q)|^{2}\;,\nonumber\\
 \int   K(q,p)\,dq\, =\,\frac{1}{2\pi\hbar}\;
|\widetilde{\Psi}(p)|^{2}\;,\label{kweew}
\end{eqnarray}
which is a fundamental requirement for   a joint distribution
in phase space. Unlike the Wigner distribution, the K-R function
is a complex function, and its operator form is not hermitian.
Nevertheless, for an arbitrary quantum  state the knowledge of the
K-R distribution at every point of the phase space allows a full
reconstruction of  this state (see \cite{K-R}).

\section{Hydrogen atom in phase space}
The definition of the K-R distribution (\ref{zrih}) allows 
one to evaluate analytical formulas of this distribution for many
complicated quantum mechanical states, for which the analytical
expression of the Wigner function is not accessible.

In the following we will present analytical results obtained for
states of the hydrogen atom. The wave functions of a
non-relativistic hydrogen atom are well known, both in position
and momentum representation. The Schr\"odinger equation with
Coulomb potential can be separated in the spherical polar
coordinates (see, e.g. \cite{B-S}). The solutions of angular part
of the equation are given by the spherical harmonics
$Y_{lm}(\theta,\varphi)$, and of the radial part can be expressed
in terms of  Laguerre functions $L^k_m$ (in the position
representation):
\begin{equation}
R_{nl}(r)= -\left[\frac{(n-l-1)!}{(n+l)!^3
(2n)}\right]^{\frac{1}{2}} \left(\frac{2
Z}{n}\right)^{\frac{3}{2}}\, e^{-\frac{Z r}{n}} \left(\frac{2 Z
r}{n}\right)^l L^{2l+1}_{n+l}\left(\frac{2 Z r}{n}\right),
\label{lag}
\end{equation}
 or  Gegenbauer functions  $C^k_m$  (in the momentum representation):
\begin{equation}
F_{nl}(p)=\left[\frac{2}{\pi}\frac{(n-l-1)!}{(n+l)!}
\right]^{\frac{1}{2}} n^2\, 2^{2(l+1)}\, l!\, \frac{n^l p^l}{(n^2
p^2+1)^{l+2}}\, C^{l+1}_{n-l-1} \left(\frac{n^2 p^2-1}{n^2 p^2+1}
\right). \label{geg}
\end{equation}
As usual, $n$, $l$, $m$ denote principal,  orbital,
and magnetic quantum numbers, respectively.

Substituting the expressions (\ref{lag}, \ref{geg})  into the
definition of  the K-R distribution Eq. (\ref{zrih}), and inserting the angular
relation, we obtain the general formulas for K-R functions of the
hydrogen atom states. In the next Section we shall present
graphical results obtained for selected hydrogen quantum numbers.
 It is worth noting that the absolute square of the
K-R function:
\begin{equation}
  \mid K(r,\theta,\varphi,p,\theta',\varphi')\mid^2\,=  \frac{1}{(2\pi)^6}\, | R_{nl}(r)
Y_{lm}(\theta,\varphi)|^{2}\; | F_{nl}(p) Y_{lm}(\theta',\varphi')
|^{2}.\nonumber
\end{equation}
is proportional to the product of probability densities in
position and momentum representations ($\hbar=1$). Thus, one can
treat it as the cornerstone of the phase space analysis of
physical systems.

\section{Examples}
Figures 1 to 5 present  phase space  K-R representation of $1s$, $2s$, $2p$, $10m$, and $10l$ states of the hydrogen atom.
Let us note that  for a 3-dimensional systems the K--R
distribution is a function of 6 variables
$(r,\theta,\varphi,p,\theta',\varphi')$, so a graphical phase
space representation can only present selected cross-sections.
We have decided to use the following method to represent the hydrogen
atom in phase space:  all the  Figures
labelled by (a), (b) (later called Fig. X.a, X.b) are  auxiliary ones -- they
present  hydrogen atom wave functions in
position and momentum representation and  the  arrows depicted  on
those figures  show the directions chosen for the
cross-sections. The corresponding cross-sections of the K-R phase space functions  are plotted  in
Figures labelled by (c) and (d) (Fig. X.c, X.d).

All the Figures are organized as follows: Fig
X.a shows cross-section $\varphi=const$ of the absolute square of hydrogen atom wave function in the
position representation, i.e.  $\mid R_{nl}(r)
Y_{lm}(\theta,\varphi)\mid^2$ which is the probability density of
finding an electron at point
$(r,\theta,\varphi)$  (and which does not depend on the value of $\varphi$).
Fig. X.b shows  cross-section $\varphi'=const$ of  the absolute square of the same wave function in the
momentum representation, i.e.  $\mid F_{nl}(p) Y_{lm}(\theta',\varphi')\mid^2$
which is the probability density
of finding an electron with momentum $(p,\theta',\varphi')$ (and here again,
a cross-section  does not depend on the
value of $\varphi'$).
 The  arrows depicted on the  Fig. X.a, X.b
 show the directions chosen for the
cross-sections of the corresponding K-R phase space functions that
are presented in Fig. X.c, where the real part of the K--R
distribution and appropriate contour plot are shown,  and in Fig
X.d, where the absolute value of this K--R distribution and its
contour plot are plotted.

Let us first study the K-R function of the $  1s$ state.
Well-known solutions of the radial part of the Schr\"odinger
equation are given by, in atomic unit:
\begin{eqnarray}
R_{10}(r)=&2 e^{-r}\nonumber &\quad \text{(in the position representation), and}\\
F_{10}(p)=&\sqrt{\frac{2}{\pi}}\frac{4}{(1+p^2)^2}\nonumber &\quad
\text{(in the momentum representation).}
\end{eqnarray}
Fig. 1.a shows spatial probability density $\mid
R_{10}(r)/(4\pi)\mid^2$ of finding   an electron at certain
point, and Fig. 1.b shows momentum probability density $\mid
F_{10}(p)/(4\pi)\mid^2$ of finding   an  electron with
definite momentum. Note, that those are not radial distributions
and they have nonzero values at $r=0$ and $p=0$ as the wave
functions were not multiplied by $r^2$ or $p^2$ factors. The K-R
distribution of $1s$ state is given by:
\begin{eqnarray}
K_{10}(r,\theta,\varphi,p,\theta',\varphi')=(2\pi^3)^{-\frac{3}{2}}
\frac{e^{-r}}{(1+p^2)^2}\exp\bigl(-i p r \cos\Theta\bigr), \label{s1}
\end{eqnarray}
where $\Theta$ denotes   the angle between $r$ and $p$:
\begin{equation}
\cos\Theta= \cos(\theta - \theta') + \bigl(\cos(\varphi - \varphi')-1 \bigr)
\sin\theta \sin\theta'.
\label{q}
\end{equation}
Fig. 1.c shows  the real part of Eq. (\ref{s1}), $K_{{\mathrm{Re}}\;1s}$,
multiplied by $(2\pi)^3$:
\begin{equation}
K_{{\mathrm{Re}}\;1s}=(2\pi)^3
{\mathrm{Re}}\biggl[K_{10}\biggl(r,\frac{\pi}{2},0,p,\frac{\pi}{2},0\biggr)\biggr].
\end{equation}
The cross-section is chosen in the directions depicted by arrows
in Fig. 1.a, 1.b. Next   we have displayed its contour plot.
The absolute value of  Eq. (\ref{s1})
multiplied by $(2\pi)^3$  is:
\begin{equation}
K_{{\mathrm{Abs}}\;1s}=(2\pi)^3{\mathrm{Abs}}
\biggl[K_{10}\biggl(r,\frac{\pi}{2},0,p,\frac{\pi}{2},0\biggr)\biggr]
\end{equation}
 and its contour plot  is presented in Fig
1.d. In this case, $K_{{\mathrm{Re}}}$ and $K_{{\mathrm{Abs}}}$ look 
quite similar mainly because  the wave function both in position and
momentum representation   decreases rapidly with the increase
of $r$ or $p$, and the oscillating  $\cos(rp)$-like structure
characteristic for the real part of K-R distribution is not
clearly seen   on the scale used to make   these
figures.   These oscillations are merely marked in the
contour plot of $K_{{\mathrm{Re}}\;1s}$ by dashed lines which correspond to
$K_{{\mathrm{Re}}}=0$.

Next we shall examine the K--R distribution of 2s and 
2p states. Solutions of the radial part of the
Schr\"odinger equation in momentum representation are given by:
\begin{eqnarray}
F_{20}(p)&=&\frac{32}{\sqrt{\pi}}\frac{4p^2-1}{(1+4p^2)^2}, \nonumber\\
F_{21}(p)&=&\frac{128}{\sqrt{3\pi}}\frac{p}{(1+4p^2)^2}.\nonumber
\end{eqnarray}
Thus, the corresponding K--R distribution are of the form:
\begin{eqnarray}
&K_{20}(r,\theta,\varphi,p,\theta',\varphi')&=\sqrt{(2^5\pi^9)}
(2-r) {e^{-\frac{r}{2}}}\frac{4p^2-1}{(1+4p^2)^2}\exp\bigl(-i p r
\cos\Theta\bigr),
\label{s2}\\
&K_{21}(r,\theta,\varphi,p,\theta',\varphi')&=\frac{\sqrt{(2\pi^9)}}{3}
r {e^{-\frac{r}{2}}}\frac{p}{(1+4p^2)^2}\exp\bigl(-i p r
\cos\Theta\bigr) \cos\theta\cos\theta', \label{p2}
\end{eqnarray}
where $\Theta$ is defined by Eq. (\ref{q}). The real part of Eq.
(\ref{s2}) and appropriate contour plot are shown   in Fig.
2.c and the real part of Eq. (\ref{p2}) in Fig. 3.c. The absolute
value of Eqs. (\ref{s2},\ref{p2}) and their contour plots are
presented in Fig. 2.d  and Fig. 3.d, respectively.  The
plot presented in Fig. 3.d has one maximum at
$(r,p)=(2,\sqrt{3}/6)$  the location of which does not
depend on variables $\theta$, $ \varphi$, $\theta'$, $\varphi'$
chosen for the cross section. This result is in perfect agreement
with what we expect for the state which   has quantum numbers
$n$, $l$   satisfying the condition $n-l=1$. In Fig. 3.c,
where the real part of   the K--R distribution is presented,
we see   the additional minimum. It is due to $\cos(rp)$-like
oscillations of the  real part of   the K--R distribution.
Actually, there is an infinite number of alternate maxima and
minima but their amplitude   decreases
 rapidly and only two are
seen   on this scale. Dashed lines depicted on this (and
every other contour-plot) denote $K_{Re}=0$. According to the
results of Fig. 2.d the absolute value of   the K--R
distribution for   the $2s$ state also has quite   a
simple form: we notice one global and three local maxima
on this cross section.
 Fig. 2.c,   where the real part of the same function is
presented, is much more complicated when we look at it   close
up (see contour plot):  Extrema are localized on lines where
$\cos(rp)$   achieves its extremal value, but on this regular
structure is  another one --   the wave function dependence
on $r$ and $p$ that is responsible for   the appearance of
additional zeros and sign--changes of   the plotted figure.

Finally we shall present an instructive example of a Rydberg
state.
Figure \ref{f4} shows the K-R distribution for a
state with $n=10$ and $l=9$.  The real part of   the
K--R distribution   is shown in Fig. \ref{f4}.c, and the
absolute value   is shown in Fig. \ref{f4}.d. Just as one
would expect there is one peak in the absolute value plot and
similar, but modified by $\cos(rp)$, structure in the real part 
of the plot. Comparing   the figures with that shown
in Fig. \ref{f3}.c-d we notice similarities in their structure.
Obviously, peaks are located at different values of $(r,p)$ but
there   is the property that for states with $n-l=1$ we
observe one maximum in the absolute value plot. Its location
changes with the increase of $n$ (it moves towards larger $r$ and
smaller $p$) and   the maximum  value gets much smaller, but
it is apparent this class of states  has much in common.
Figure \ref{f5} shows the K--R distribution for state $n=10$,
$l=8$. Analysis similar to   that made above holds true when
we compare Fig. \ref{f5} and \ref{f2}. Without counting that all
the scales have changed we find 4 extrema ($(n-l)^2$) in the
absolute value plots and similar oscillating structures in the
real part plots. It is  a general and well known result that
$n-l$ is equal to the number of extrema of hydrogen atom wave
functions both in position and momentum representation. We only
stress that the K-R distribution renders very well this property:
its absolute value has $(n-l)^2$ extrema.

\section{Summary remarks}
We have presented   a phase space representation of the
hydrogen atom using   the K-R distribution. An advantage of
such  a representation is that the system can be treated
analytically. It is an attractive feature of the K-R distribution
that as long as one knows   the wave function of the system
there is no need to perform any integrals to obtain this
distribution value in every point of phase space.

We have described in detail   the  $1s$, $2s$, $2p$, $10m$,
and $10l$ states of the hydrogen atom. Plots corresponding to 
the  K--R functions of these states were presented and compared
with the plots of probability densities in position and momentum
representations.

\section*{Acknowledgments}

We thank S. Daffer for comments about the final draft. This work
was partially supported by a KBN grant 2P03 B 02123, and the
European Commission through the Research Training Network QUEST.

\newpage

\begin{figure}

\caption{ The K-R phase space representation of $1s$ state ($n=1$, $l=0$):\\
(a) the cross section
$\varphi=const$ of spacial probability density $\mid R_{10}(r)/(4\pi)\mid^2$
of finding electron at point $(r,\theta,\varphi)$; \\
(b)  the cross section $\varphi'=const$ of momentum probability density
$\mid F_{10}(p)/(4\pi)\mid^2$ of finding electron with momentum
$(p,\theta',\varphi')$;\\
(c) the real part of the K-R distribution and its contour plot (the cross section
is made in directions shown by arrows depicted on  Fig. 1(a) and (b));\\
(d) the absolute value of the K-R distribution and its contour plot
(the same  cross section as above).
 }
\label{f1}
\end{figure}
\begin{figure}

\caption{The K-R phase space representation of $2s$ state ($n=2$,
$l=0$):\\
(a) the cross section
$\varphi=const$ of spacial probability density $\mid R_{20}(r)/(4\pi)\mid^2$
of finding electron at point $(r,\theta,\varphi)$; \\
(b)  the cross section $\varphi'=const$ of momentum probability density
$\mid F_{20}(p)/(4\pi)\mid^2$ of finding electron with momentum
$(p,\theta',\varphi')$;\\
(c) the real part of the K-R distribution and its contour plot (the cross section
is made in directions shown by arrows depicted on  Fig. 1(a) and (b));\\
(d) the absolute value of the K-R distribution and its contour plot
(the same  cross section as above).
}
\label{f2}
\end{figure}
\begin{figure}
\caption{ The K-R phase space representation of $2p$ state ( $n=2$,
$l=1$
(a) the cross section
$\varphi=const$ of spacial probability density $\mid R_{21}(r)/(4\pi)\mid^2$
of finding electron at point $(r,\theta,\varphi)$; \\
(b)  the cross section $\varphi'=const$ of momentum probability density
$\mid F_{21}(p)/(4\pi)\mid^2$ of finding electron with momentum
$(p,\theta',\varphi')$;\\
(c) the real part of the K-R distribution and its contour plot (the cross section
is made in directions shown by arrows depicted on  Fig. 1(a) and (b));\\
(d) the absolute value of the K-R distribution and its contour plot
(the same  cross section as above).}
\label{f3}
\end{figure}
\begin{figure}

\caption{  The K-R phase space representation of $10m$ state ($n=10$,
$l=9$):\\
(a) the cross section
$\varphi=const$ of spacial probability density of finding electron at point $(r,\theta,\varphi)$; \\
(b)  the cross section $\varphi'=const$ of momentum probability density of finding electron with momentum
$(p,\theta',\varphi')$;\\
(c) the real part of the K-R distribution and its contour plot (the cross section
is made in directions shown by arrows depicted on  Fig. 1(a) and (b));\\
(d) the absolute value of the K-R distribution and its contour plot
(the same  cross section as above).}
\label{f4}
\end{figure}
\begin{figure}

 \caption{  The K-R phase space representation of $10l$ state ($n=10$, $l=8$):\\
(a) the cross section
$\varphi=const$ of spacial probability density of finding electron at point $(r,\theta,\varphi)$; \\
(b)  the cross section $\varphi'=const$ of momentum probability density of finding electron with momentum
$(p,\theta',\varphi')$;\\
(c) the real part of the K-R distribution and its contour plot (the cross section
is made in directions shown by arrows depicted on  Fig. 1(a) and (b));\\
(d) the absolute value of the K-R distribution and its contour plot
(the same  cross section as above).}
 \label{f5}
 \end{figure}

\end{document}